\documentclass[sigconf]{acmart}
\usepackage{amsmath,amsfonts}
\usepackage{algorithmic}
\usepackage{graphicx}
\usepackage{textcomp}
\usepackage{xcolor}
\usepackage{booktabs}
\usepackage{subcaption}

\def\BibTeX{{\rm B\kern-.05em{\sc i\kern-.025em b}\kern-.08em
    T\kern-.1667em\lower.7ex\hbox{E}\kern-.125emX}}

\usepackage{multirow}

\def\tablehack{\vspace*{-10pt}}

\newcommand{\mypara}[1]{\smallskip\noindent\textit{#1}\quad}

\def\rqOne{RQ1: Can partial test oracles precisely detect BCI faults?}
\def\rqTwo{RQ2: Can corrective heuristics accurately transform faulty model outputs?}
\def\rqThree{RQ3: Can BCI faults be localized to input and output slices?}

\def\rqFour{RQ4: Can fault-based data acquisition and model retraining repair BCI faults?}
\def\rqFive{RQ5: Are corrective heuristics or fault localization sufficient on their own?}

\def\tableHeuristics{%
\begin{table*}[t]
\caption{\textbf{Corrective Heuristics for BCI Faults.}}
\begin{center}
\tablehack
\begin{tabular}{p{0.19\linewidth}  p{0.15\linewidth}  p{0.62\linewidth}}
\toprule
\textbf{Fault} & \textbf{\mbox{BCI Application(s)*}}
& \textbf{Description of Corrective Heuristic} \\
\midrule
\multirow{2}{*}{}Temporal Inconsistency & OE, SS & Select the most common state predicted by the decoder on surrounding data points \\
& CC, FN & Compute the average of decoder predictions on surrounding data points  \\
\addlinespace
\multirow{2}{*}{}Illegal Transition & MD & Sample from state transition probabilities learned on training dataset  \\
& SS & Select a state predicted by decoder on previous data point \\
\addlinespace
Rapid Motion & FN & Compute the mean of the most recent outputs or closest possible output \\
\addlinespace
Multimodal Inconsistency & SS & Compute the most common state predicted by the decoder on previous data points \\
\addlinespace
Out of Bounds & FN & Compute the mean of the most recent outputs or the closest in-bounds value \\
\bottomrule
\\[-10pt]
\multicolumn{3}{l}{\small{*MD: Motor Decoding, OE: Observe-or-Execute Classification, SS: Sleep Stage Classification, CC: Cursor Control, FN: Field Navigation.}}
\end{tabular}
\label{table:faults}
\end{center}
\end{table*}}

\def\tablePrevalence{%
\begin{table*}[t]
\caption{\textbf{Frequency of BCI Faults Detected by Partial Test Oracles.\label{table:prevalence}}
Precision is the percentage of correctly detected faulty outputs, given ground truth. $^{*}$For continuous-output applications the precision is generally unknowable without defining an epsilon up to which model outputs are considered equal; for some faults such as
out of bounds (OB), however, precision is guaranteed to be 100\%.}
\begin{center}
\tablehack
\begin{tabular}{p{0.3\linewidth}  p{0.24\linewidth}  p{0.13\linewidth}  p{0.15\linewidth}}
\toprule
\textbf{BCI Application} & \textbf{Fault}&\textbf{Precision} &\textbf{Frequency} \\
\midrule
\multirow{1}{*}{Motor Decoding (MD)}& Temporal Inconsistency (TI) & $76\%$ &$12.7\%$   \\
 & Illegal Transition (IT) & $91\%$ &$3.13\%$ \\
\midrule
\multirow{1}{*}{Observe-or-Execute Classification (OE)}&Temporal Inconsistency (TI)& $62\%$ & $37.63\%$ \\\midrule
Sleep Stage Classification (SS) & Temporal Inconsistency (TI) & $59\%$& $26.6\%$  \\ 
& Illegal Transition (IT) & $72\%$ & $7.66\%$ \\ 
& Multimodal Inconsistency (MM) & $94\%$ & $7.35\%$  \\ \midrule
\multirow{1}{*}{Cursor Control (CC)}& Temporal Inconsistency (TI) & ---$^{*}$& $0.69\%$\\
& Rapid Motion (RM) & ---$^{*}$& $0.26\%$\\\midrule
Field Navigation (FN)& Temporal Inconsistency (TI) & ---$^{*}$& $7.89\%$\\
 & Rapid Motion (RM) & ---$^{*}$& $7.89\%$ \\
& Out of Bounds (OB) & 100\%$^{*}$& $1.28\%$ \\
\bottomrule
\end{tabular}
\label{table:prevalence_table}
\end{center}
\vspace*{-8pt}
\end{table*}}

\def\tableRepair{%
\begin{table*}
\caption{\textbf{Repairing BCIs with Fault-Based Data Acquisition.} For each application, we seek to eliminate either a single fault or all faults combined. We compare the number of faults (averaged over 10 trials) before retraining (Baseline), after retraining with data acquired based on the natural 
distribution of tasks (After Bernoulli), and after retraining with data acquired based on a fault-based distribution of tasks (After Fault Localization w/ Corrections).
We also compare to simply applying corrective heuristics (After Corrections) and fault-based data acquisition without applying corrective heuristics first (After Fault Localization w/o Corrections).
Comparison to baseline: $\textsuperscript{***}p<0.01$, $ \textsuperscript{**}p < 0.05$, $\textsuperscript{*}p < 0.1$.
Comparison to Bernoulli-sampled task distribution: $\textsuperscript{+++}p< 0.01$, $\textsuperscript{++} p < 0.05$, $\textsuperscript{+}p < 0.1$}
\begin{center}
\tablehack
\setlength{\aboverulesep}{0pt}
\setlength{\belowrulesep}{0pt}
\begin{tabular}{p{0.19\linewidth}  p{0.19\linewidth}  p{0.07\linewidth} p{0.08\linewidth}  p{0.12\linewidth} | p{0.10\linewidth} p{0.12\linewidth}  }
\toprule
    \textbf{BCI Application} & \textbf{Fault} & \textbf{Baseline} & \textbf{\begin{tabular}[c]{@{}l@{}}After  \\ Bernoulli\end{tabular}}& \textbf{\begin{tabular}[c]{@{}l@{}}After Fault \\ Localization \\ w/ Corrections\end{tabular}} & 
\textbf{\begin{tabular}[c]{@{}l@{}}After  \\ Corrections\end{tabular}}   & \textbf{\begin{tabular}[c]{@{}l@{}}After Fault \\ Localization \\ w/o Corrections\end{tabular}}
\\ \midrule
EEG Motor Decoding& Temporal Inconsistency & 230.75 & 241.96  & 244.79 $^{**}$ & -  & 240.17 $^{**}$
\\
& Illegal Transition & 78.29 & 76.38  & 73.96 & -  & 207.79 $^{***+++}$
\\
& All faults combined & 309.04 & 318.33 & 308.13  & 266.25 $^{***+++}$  & 305.88 
\\ \midrule
Observation-or-Execution & Temporal Inconsistency & 62.56 & 61.44  & 46.22 $^{***++}$ & - & 53.78 
\\
Classification& All faults combined & 62.56 & 61.44  & 46.22 $^{***++}$ & 71.67 $^{*}$  & 53.78 \\ \midrule
\multirow{3}{*}{\begin{tabular}[c]{@{}l@{}}Sleep Stage \\ Classification \end{tabular}}& Temporal Inconsistency & 153.00 & 147.04 & 151.67 $^{+++}$  & -  & 153.83 $^{+}$
\\
& Illegal Transition & 31.92 & 27.50  & 44.50 $^{***+++}$ & - & 24.38 $^{**++}$
\\
& Multimodal Inconsistency & 23.79 & 19.79  & 25.83 $^{++}$ & -  & 51.75 $^{***+++}$
\\
& All faults combined & 208.71 & 194.33   & 199.88 $^{*}$ & 221.17 $^{**+++}$ & 199.25 $^{**}$\\
 \midrule
BCI Cursor Control& Temporal Inconsistency & 100.00 & 63.00  & 62.00 $^{*}$ & -  & 62.00 
\\
& Rapid Motion & 349.83 & 296.50  & 292.50  & - & 293.17 
\\
& All faults combined & 449.83 & 359.50  & 355.67 $^{*}$  & 525.50 $^{+++}$ & 358.50 $^{*}$\\
 \midrule
Field Navigation& Temporal Inconsistency & 1220.20 & 1200.40  & 1178.40 $^{**}$ & - & 1161.40 $^{**++}$
\\
& Rapid Motion & 2002.00 & 2048.20 & 2040.60 & -  & 2001.40 
\\
& Out of Bounds & 93.60 & 77.40  & 69.80 & - & 84.80 $^{+}$
\\
& All faults combined & 3315.80 & 3326.00 & 3274.60  & 3305.20  & 3282.60 \\
\bottomrule
\end{tabular}
\label{table:repair_loc}
\end{center}
\tablehack
\end{table*}}

\def\tableTaxonomy{%
\begin{table*}
\caption{\textbf{Classification of BCI Faults.}}
\begin{center}
\tablehack
\setlength{\tabcolsep}{4pt}
\resizebox{\linewidth}{!}{%
\begin{tabular}{p{0.15\linewidth}  p{0.23\linewidth}  p{0.28\linewidth}  p{0.23\linewidth}  p{0.16\linewidth}}
\toprule
\textbf{Category} & \textbf{Fault}& \textbf{Description}& \textbf{Implication}& \textbf{BCI Application(s)*} \\
\midrule
Input Validation & Artifact (A) & Signal noise from irrelevant motor or eye movement & Interference with neural signal input to decoder & --- \\
\midrule
\multirow{3}{*}{}Temporal Validation & Temporal Inconsistency (TI) & Flickering in decoder predictions& Undesirable behavior& MD, OE, CC, FN, SS  \\
\addlinespace
& Illegal Transition (IT) & Impossible/improbable transition between predicted states&Impossible/undesirable behavior& MD, SS  \\
\addlinespace
& Rapid Motion (RM) & Improbably large change in prediction&Undesirable/unsafe behavior& CC, FD  \\
\midrule
Consistency & Multimodal Inconsistency~(MM) & Disagreement between predictions by decoder and by auxiliary model & Potentially incorrect prediction & SS \\
\midrule
Domain Knowledge & Out of Bounds (OB) & Impossible prediction & Impossible behavior & CC \\
\bottomrule
\\[-10pt]
\multicolumn{5}{l}{\small{*Evaluated BCI Applications (Table~\ref{table:decoders}). MD: Motor Decoding, OE: Observe-or-Execute Classification, SS: Sleep Stage Classification, CC: Cursor Control, FN: Field Navigation.}}
\end{tabular}}
\label{table:fault_taxonomization}
\end{center}
\vspace*{-8pt}
\end{table*}}

\def\tableApps{%
\begin{table*}[t]
\caption{\textbf{Summary of BCI Applications Used for Evaluation.}}
\tablehack
\begin{center}
\setlength{\tabcolsep}{3pt}
\resizebox{\linewidth}{!}{%
\begin{tabular}{lllllll}
\hline
\toprule
\textbf{\begin{tabular}[c]{@{}l@{}}BCI  \\ Application\end{tabular}} & \textbf{\begin{tabular}[c]{@{}l@{}}Input  \\ Modality\end{tabular}} & \textbf{\begin{tabular}[c]{@{}l@{}}Output  \\ Prediction\end{tabular}} & \textbf{\begin{tabular}[c]{@{}l@{}}Control  \\ Mechanism\end{tabular}} & 
\textbf{\begin{tabular}[c]{@{}l@{}}Intended  \\ Use\end{tabular}} & 
\textbf{\begin{tabular}[c]{@{}l@{}}Decoder  \\ Model*\end{tabular}} & 
\textbf{\begin{tabular}[c]{@{}l@{}}Performance  \\ Metric\end{tabular}} \\
\midrule
Motor Decoding (MD)& EEG & One of 3 hand postures& Active& Restorative &LSTM& Classification Accuracy  \\ 
Observe-or-Execute Classification (OE)& EEG & Observation or execution& Active & Restorative & SVM & Classification Accuracy \\ 
Sleep Stage Classification (SS)& EEG& One of four sleep stages& Passive & Diagnostic & RF & Classification Accuracy  \\ 
Cursor Control (CC)& SNR & 2D cursor position& Active & Restorative & WCD & Mean Squared Error  \\ 
Field Navigation (FN)& SNR & 2D position in field & Active & Diagnostic & Neural Net & Mean Squared Error  \\
\bottomrule
\\[-10pt]
\multicolumn{7}{l}{\small{*SNR: Single neuron recordings; SVM: Support vector machine; RF: Random forest classifier; WCD: Weiner Cascade Decoder.}}
\end{tabular}}
\label{table:decoders}
\end{center}
\vspace*{-8pt}
\end{table*}}

\def\figureHeuristics{%}
\begin{figure}
    \vspace{-1.0cm}
    \centerline{\includegraphics[scale=0.28,angle=270,origin=c,clip,trim={0 0 2.1cm 0}]{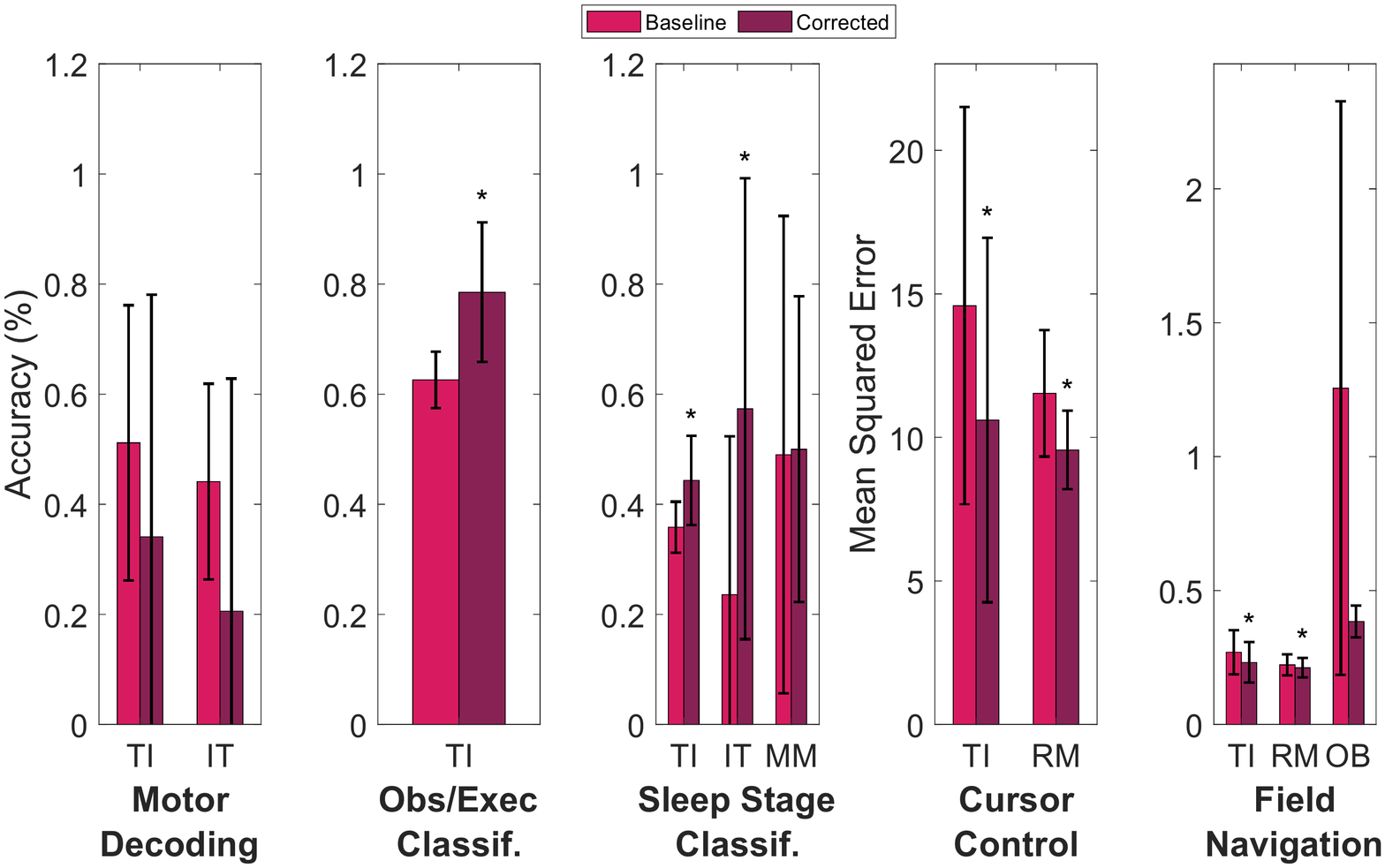}}
    \vspace{-1.4cm}
    \caption{\textbf{Efficacy of Corrective Heuristics.} An asterisk (*) indicates a significant difference (paired two-sample t-test, p-value$<$0.05). Note that for mean squared error, lower is better.\label{fig:decoder_correction}}
\end{figure}}

\def\figureLocalization{%}
\begin{figure}
    \begin{subfigure}[t]{\linewidth}
    \vspace{-1.1cm}
    \includegraphics[angle=-90,scale=0.28]{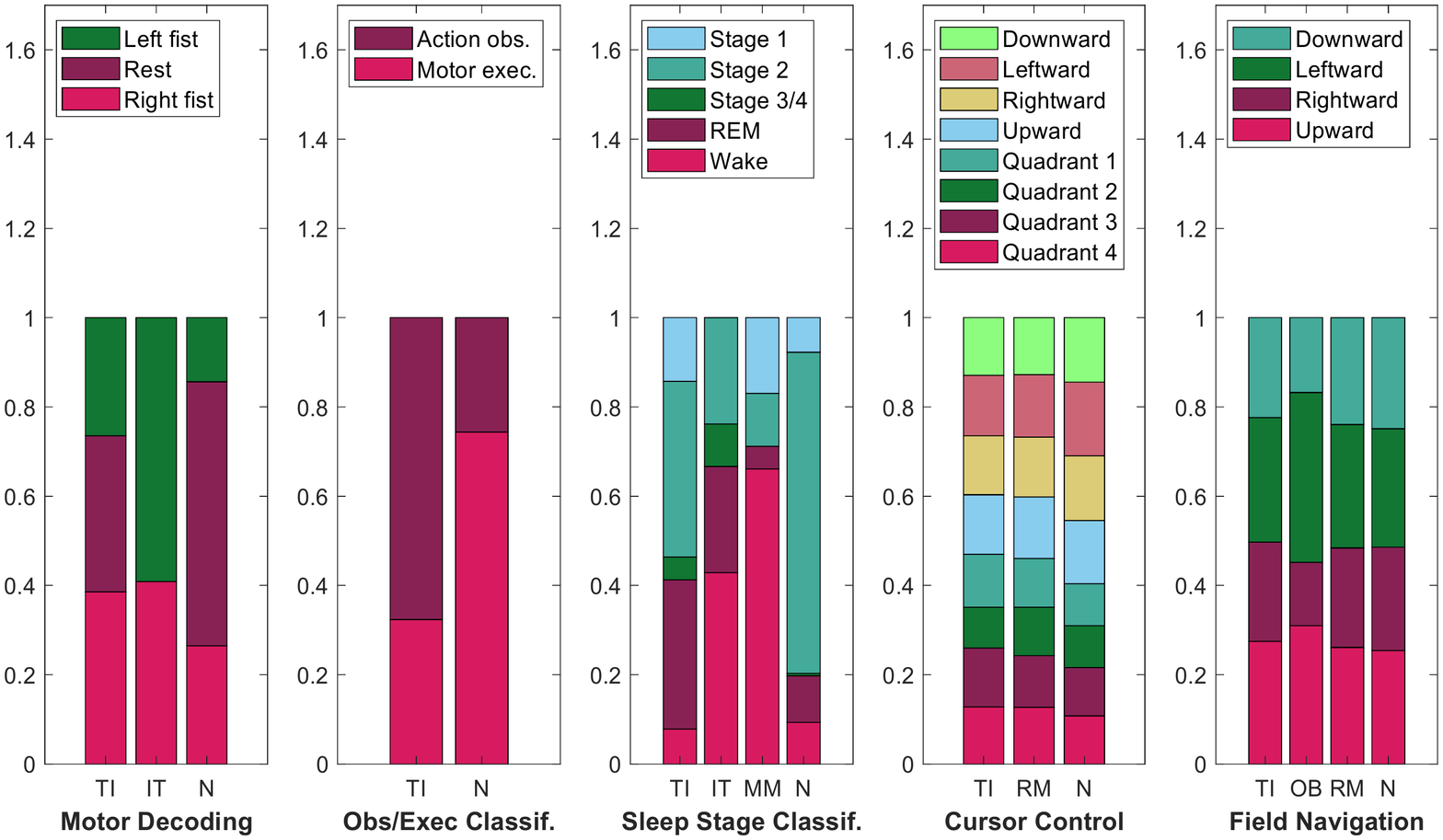}
    \vspace{-9mm}
    \caption{\textbf{Output Slices.}}
    \label{fig:fault_task_loc}
    \end{subfigure}
    \begin{subfigure}[t]{\linewidth}
    \includegraphics[angle=-90,scale=0.28,trim=90 0 0 0, clip]{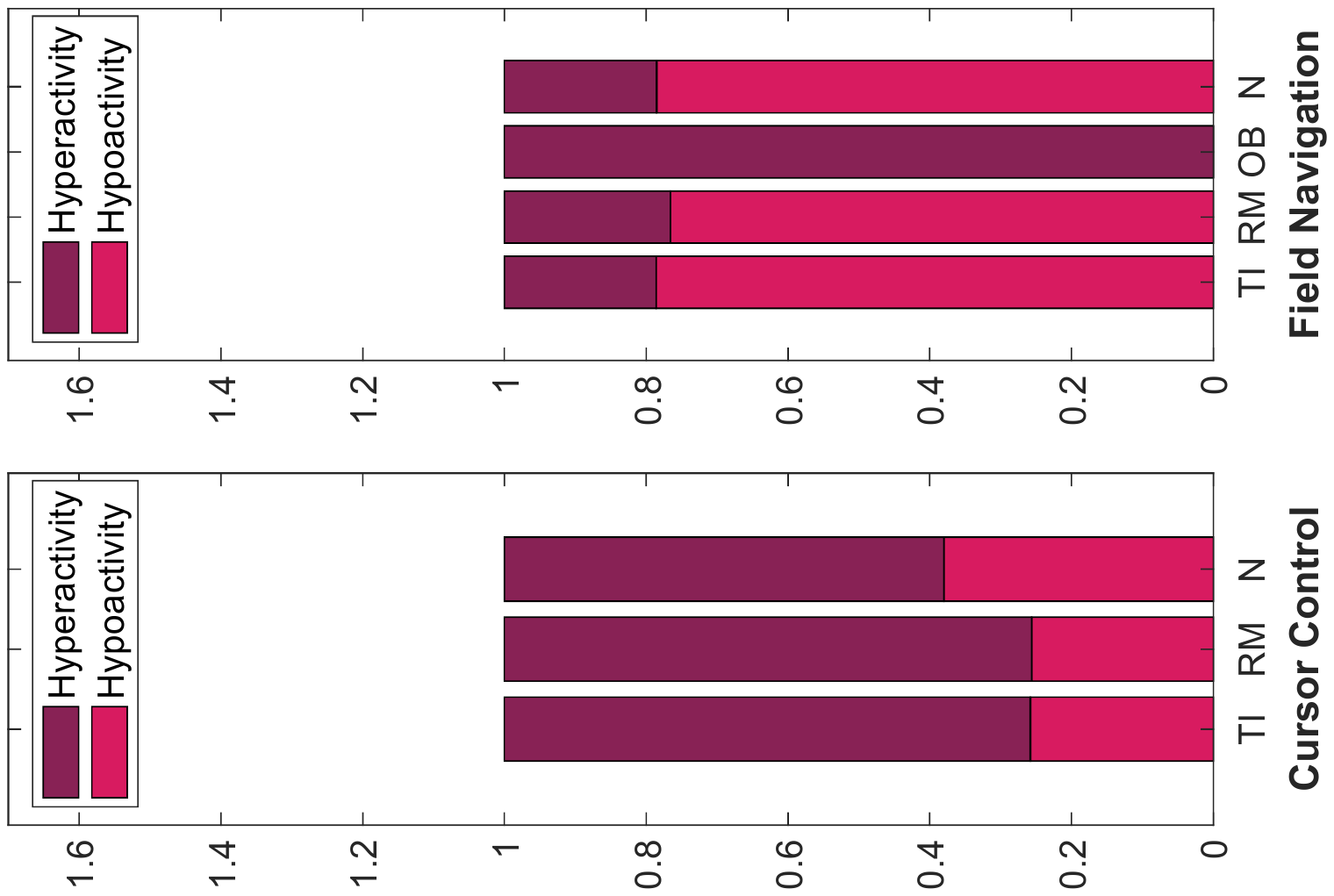}
    \vspace{-1cm}
    \caption{\textbf{Input Slices.}}
    \label{fig:fault_task_barplot}
    \end{subfigure}
    \vspace*{-10pt}
    \caption{\textbf{Localizing faults to specific output slices (top) and input slices (bottom).}
    Different faults (on the x-axis, with N referring to no fault) exhibit different distributions of co-occurring slices. All distributions are normalized to 1, but slices may co-occur.}
    \label{fig:fault_task}
\end{figure}}

\AtBeginDocument{%
  \providecommand\BibTeX{{%
    \normalfont B\kern-0.5em{\scshape i\kern-0.25em b}\kern-0.8em\TeX}}}

\copyrightyear{2022}
\acmYear{2022}
\setcopyright{acmcopyright}\acmConference[ICSE '22]{44th International Conference on Software Engineering}{May 21--29, 2022}{Pittsburgh, PA, USA}
\acmBooktitle{44th International Conference on Software Engineering (ICSE '22), May 21--29, 2022, Pittsburgh, PA, USA}
\acmPrice{15.00}
\acmDOI{10.1145/3510003.3512764}
\acmISBN{978-1-4503-9221-1/22/05}
\setcopyright{none}

\acmConference[ICSE 2022]{The 44th International Conference on Software Engineering}{May 21–29, 2022}{Pittsburgh, PA, USA}

\begin{document}

\title[Repairing Brain-Computer Interfaces with Fault-Based Data Acquisition]{Repairing Brain-Computer Interfaces\\with Fault-Based Data Acquisition}

\author{Cailin Winston}
\affiliation{
  \institution{University of Washington}
  \city{Seattle}
  \state{Washington}
  \country{USA}
  \postcode{98195}
}
\email{cailinw@cs.washington.edu}

\author{Caleb Winston}
\affiliation{
  \institution{University of Washington}
  \city{Seattle}
  \state{Washington}
  \country{USA}
  \postcode{98195}
}
\email{calebwin@cs.washington.edu}

\author{Chloe N Winston}
\affiliation{
  \institution{University of Washington}
  \city{Seattle}
  \state{Washington}
  \country{USA}
  \postcode{98195}
}
\email{chloewin@cs.washington.edu}

\author{Claris Winston}
\affiliation{
  \institution{University of Washington}
  \city{Seattle}
  \state{Washington}
  \country{USA}
  \postcode{98195}
}
\email{clarisw@cs.washington.edu}

\author{Cleah Winston}  
\affiliation{
  \institution{University of Washington}
  \city{Seattle}
  \state{Washington}
  \country{USA}
  \postcode{98195}
}
\email{cleahwin@gmail.com}

\author{Rajesh P N Rao}
\affiliation{
  \institution{University of Washington}
  \city{Seattle}
  \state{Washington}
  \country{USA}
  \postcode{98195}
}
\email{rao@cs.washington.edu}

\author{Ren\'e Just}
\affiliation{
  \institution{University of Washington}
  \city{Seattle}
  \state{Washington}
  \country{USA}
  \postcode{98195}
}
\email{rjust@cs.washington.edu}

\renewcommand{\shortauthors}{Winston et al.}

\begin{abstract}
%% What are BCIs?
Brain-computer interfaces (BCIs) decode recorded neural signals from the brain and/or stimulate the brain with encoded neural signals. BCIs span both hardware and software and have a wide range of applications in restorative medicine, from restoring movement through prostheses and robotic limbs to restoring sensation and communication through spellers. BCIs also have applications in diagnostic medicine, e.g., providing clinicians with data for detecting seizures, sleep patterns, or emotions.

%% The problem statement
Despite their promise, BCIs have not yet been adopted for long-term, day-to-day use because of challenges
related to reliability and robustness, which are needed for safe operation in all scenarios. 
Ensuring safe operation currently requires hours of manual data collection and recalibration, involving
both patients and clinicians. However, data collection is not targeted at eliminating specific faults in a BCI.
%% The contributions of this paper
This paper presents a new methodology for characterizing, detecting, and localizing faults in BCIs. Specifically, it proposes partial test oracles as a method for detecting faults
and slice functions as a method for localizing faults to 
characteristic patterns in the input data or relevant tasks performed by the user. 
Through targeted data acquisition and retraining,
the proposed methodology improves the correctness of BCIs.
We evaluated the proposed methodology on five BCI applications.
The results show that the proposed methodology (1) precisely localizes faults and (2) can significantly reduce the frequency of faults through retraining based on targeted, fault-based data acquisition.
These results suggest that the proposed methodology is a promising step towards repairing faulty BCIs.

\end{abstract}

\begin{teaserfigure}
    \centering
    \includegraphics[scale=0.43]{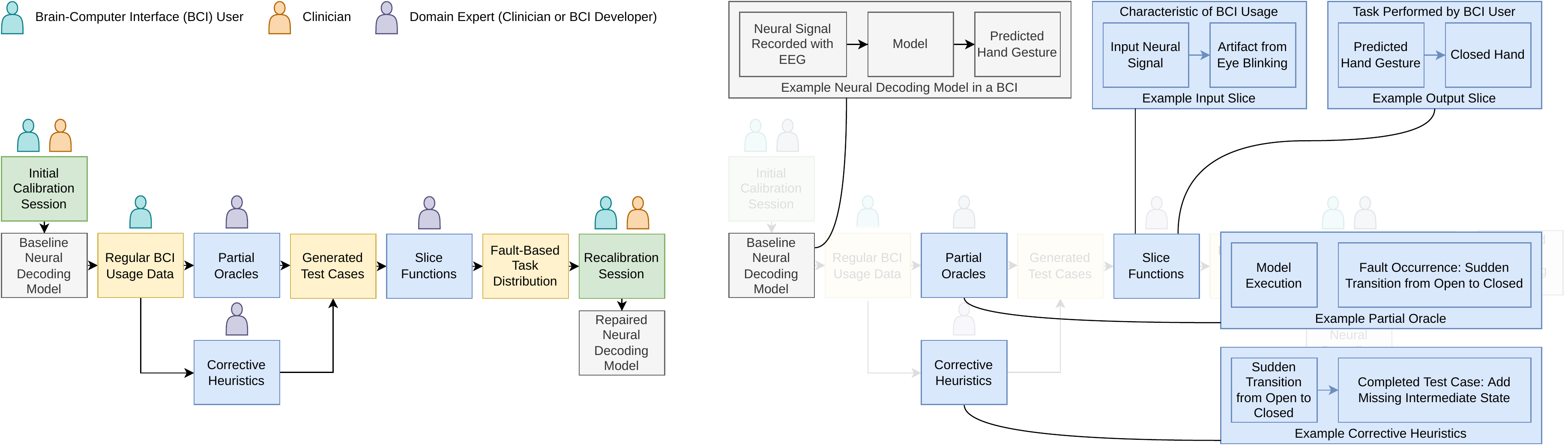}
    \caption{Proposed end-to-end methodology for repairing BCIs (left) and a concrete example for a motor decoding BCI (right).}
    \label{fig:workflow}
    %% Add a bit more space between the teaser image and the following text.
    \vspace*{30pt}
\end{teaserfigure}




\maketitle

\section{Introduction}

Brain-computer interfaces (BCIs) are systems that enable communication between the 
brain and hardware \cite{allison2007brain,rao2013bci}, by recording neural signals from the brain and/or stimulating the brain.
By decoding recorded neural signals (neural decoding) or 
encoding neural signals for stimulation (neural encoding), BCIs can restore lost function or even enhance normal 
function. For example, BCIs such as neuroprosthetic limbs have 
restored motor function in individuals with muscle impairments or spinal cord injury.
Through neural decoding, BCIs can also serve important diagnostic purposes such as
detecting seizures, sleep stages, and emotion.

Because of the close interaction between BCIs and the brain, it is critical to ensure their
safety and correctness. Unfortunately, BCIs can exhibit faulty behavior with decoded intentions (subsequently referred to as \emph{tasks})
that can be (a) impossible for a prosthetic to execute, (b) undesirable  for the user
(e.g., wavering movements), or (c) unsafe (e.g., rapid motion of the prosthetic that could
harm the user or a bystander). These are referred to as \emph{faults}. Typically, BCIs are fine-tuned during a calibration session with the
prospective user, ideally to eliminate such faults.
However, data shifts can occur during use thereafter due to the non-stationarity of brain signals and differing environmental conditions not accounted for during the controlled calibration sessions.
As a result, faults which were eliminated during initial calibration may reoccur.
To subsequently eliminate these faults,
further recalibration of the model or improved data pre-processing is required, which may involve conducting additional experiments to collect neural data while users perform tasks.
Identification of tasks a BCI performs poorly on is
solely based on user reports, which may not be comprehensive.
To the best of our knowledge,
no methodology exists for continuously testing BCIs based on a partial specification of correctness
\mbox{and subsequently localizing and repairing discovered faults.}

We propose a methodology for testing and repairing BCIs
based on expert-defined partial test oracles (partial specification of correctness), corrective heuristics (data transformation functions), and slice functions for fault localization (Figure \ref{fig:workflow}).
For a given class of BCIs, a domain expert such as
a clinician or BCI developer defines:
\begin{itemize}
    %% Better use of space
    \setlength{\itemsep}{1pt}
    \item \emph{Partial test oracles}: detect faults during BCI usage.
    \item \emph{Corrective heuristics}: transform faulty model outputs by mapping implausible output values to plausible ones.
    \item \emph{Slice functions}: slice and classify BCI inputs and outputs.
\end{itemize}
Together, these enable repairing faulty BCIs through targeted
data acquisition informed by fault localization.
Specifically, detected faults are localized to
co-occurring input or output slices.
Input slices correspond to characteristics of BCI usage, such as the individual using the BCI, environmental settings, and the quality of the pre-processed data that the underlying neural decoding model ingests.
Output slices correspond to tasks which are the various actions that a user can perform with the BCI.
Clinicians can then acquire data by targeting
tasks co-occurring with faults.
To repair the BCI, the underlying neural decoding model is retrained using the acquired data. Localizing
faults to input slices may also be used, e.g., to improve the pre-processing of the neural signal to eliminate faults stemming from artifacts in that signal.
Note that corrective heuristics transform faulty model outputs to aid localizing faults to tasks. This is distinct from repairing a BCI, which involves targeted data acquisition, based on a task distribution, and model retraining \mbox{in an attempt to reduce the likelihood of faults occurring.}

As an example, consider a BCI that predicts hand gestures
from neural signals. A domain expert may define a partial test oracle
that identifies sudden transitions between hand positions as faults,
a corrective heuristic function
that corrects a sudden-transition fault by
smoothly interpolating between the hand positions,
and slice functions that classify the various hand gestures.
If instances of such sudden-transition faults occur more
frequently for slices of the heuristically transformed output that correspond to a closed gesture (i.e., the fault is localized to the closed gesture task), clinicians and users would be recommended to collect additional data for this task. Further calibration of the gesture prediction model can reduce the number of sudden-transition faults. 
The domain expert may also define a slice function that classifies
input slices that contain artifacts from eye blinking (a 
characteristic interference in EEG data). If faults are localized
to these input slices, BCI developers could be asked to improve pre-processing to eliminate such artifacts.

We evaluated the proposed methodology on five BCI applications.
The results show that
(1) partial test oracles can detect faults in BCIs,
(2) fault localization can identify input and output slices co-occurring with faults, and
(3) targeted, fault-based data acquisition can significantly reduce the frequency of faults in BCIs.

The key contribution of this paper is an end-to-end, human-in-the-loop methodology for repairing deployed BCIs. Specific contributions include:
\begin{itemize}
    \vspace*{-3pt}
    %% Better use of space
    \item A classification of BCI applications and faults (Section~\ref{sec:classification}).
    \item A novel application of partial test oracles and fault localization to BCIs, enabling targeted fault-based data acquisition to repair faulty BCIs (Section~\ref{sec:repair_title}).
    \item A validation of the proposed methodology using five BCI applications, demonstrating that faulty BCIs can be successfully repaired post-deployment (Sections~\ref{sec:apps} and~\ref{sec:results}).
    \vspace*{-3pt}
\end{itemize}

\section{Background}
\vspace*{-3pt}

\begin{figure*}[!t]
    \centering
    \includegraphics[scale=0.4]{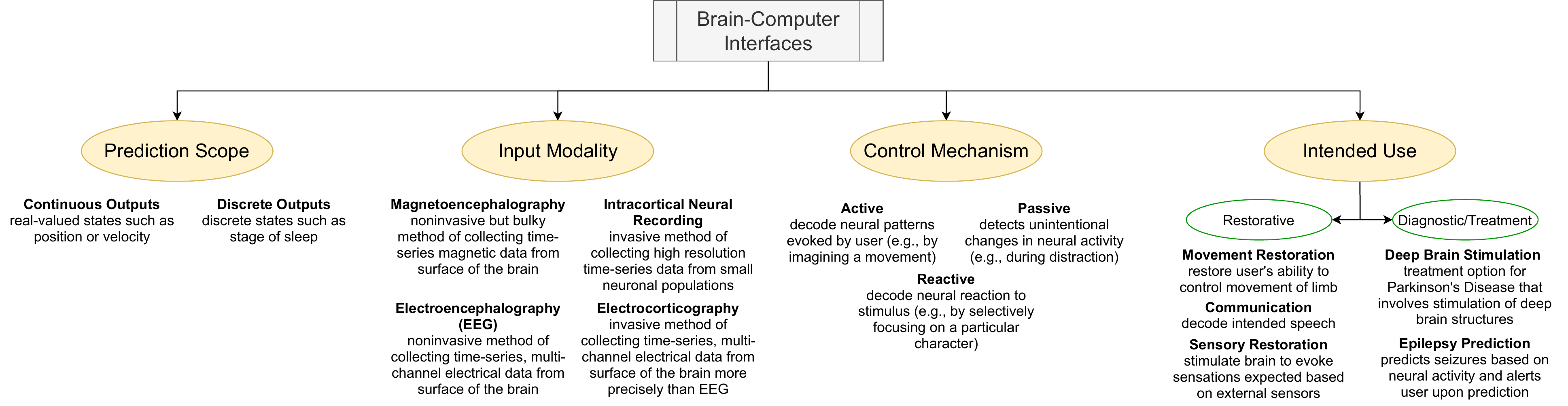}
    \vspace*{-8pt}
    \caption{\textbf{Classification of Brain-Computer Interfaces}}
    \label{fig:bcitypes}
\end{figure*}

\subsection{Brain-Computer Interfaces}

Many BCIs involve translating  neural signals, which are recorded from the brain and reflect a person's intended action (task),
into control signals (e.g., for a prosthetic limb in a paralyzed person, for an avatar in an augmented or virtual
reality environment for rehabilitation therapy, or for non-medical 
applications such as brain-controlled gaming.) A BCI has components for signal 
acquisition, signal processing and feature extraction, and translation of these signals into a meaningful output 
through neural decoding. Neural signaling in the brain utilizes electrical impulses communicated between neurons. This electrical activity can be 
measured through a variety of techniques, ranging from highly invasive procedures involving brain implants to 
noninvasive 
methods such as electroencephalography (EEG). The signal is often recorded as a 2D array, capturing
time and input channels---streams of input from electrodes placed on specific places on the scalp (e.g., EEG) or within the brain (e.g., intracortical recordings).
The recorded brain activity can be affected by artifacts, such as line noise from external electrical impulses and movement-related artifacts. The brain signal can also be weak and mixed due to signal propagation through biological tissue before measurement. Signal processing and feature extraction are typically used to reduce noise and extract task-relevant features from the data.
Even with such processing, noise often persists in the signal and can result in faults.

The processed neural data is then translated into a desired output using a neural decoder.
While some 
neural decoders are hand-crafted models with highly interpretable parameters, other decoders utilize classical
machine learning methods and, more recently, deep neural networks that can model complex patterns in data with less manual parameter tuning. The type of decoder is determined based on the application and the complexity of the neural data.

Neural decoders are often trained from hand-labeled datasets collected from studies with several participants.
However, before deploying such models to a device, such as a prosthetic, a calibration session is required with the 
individual user, due to the variability of neural signals across individuals \cite{Miralles2015}. In this session, neural data is collected while the individual performs or imagines performing
certain tasks. For example, for a hand prosthetic, an individual may be asked to move their hand in specified directions and grasp an item.
The model is calibrated on the neural data that is collected while the user performs these tasks, so that it can accurately decode the neural data and predict the true intention of the individual.

\subsection{Challenges in Deploying BCIs}
There are multiple challenges associated with deploying accurate BCIs. One challenge is the need for extensive data collection and training of 
the neural decoder. In the successful operation of any BCI, two systems are at play---the neural decoder and the user's brain. While the neural decoder learns neural patterns that correspond to specific user states, the user's brain adapts to control the BCI. Consider a motor BCI whose goal is to perform movements desired by the user. First, the neural decoder is calibrated \cite{DBLP:journals/corr/MladenovicML17}. Neural activity is recorded while the user imagines various movements (imagination is required when the individual is paralyzed or otherwise unable to physically perform the desired task) \cite{DBLP:journals/corr/MladenovicML17}. This requires hours of experiments in the clinic, with the user performing many tasks that the BCI would be expected to perform. The data must also be manually labeled and annotated in some applications, and this may be error-prone. A neural decoder is fit to this data. Next, the user begins using the BCI and learning how to control it. If faults persist, further calibration may be required.

Another challenge is that in current BCI applications, it is hard to monitor and characterize faults after deployment.
If a BCI performs poorly, a user may report failure scenarios, which a clinician may use to prescribe additional
data collection sessions. Faults that are not reported are not necessarily accounted for, and thus it is hard to
assess the performance and safety of the device.

\subsection{Software Testing and Debugging}

Recognizing the need for a methodology for identifying and repairing faults in BCIs post deployment,
we explore the applicability of existing software testing constructs to this emerging domain. Partial test oracles, which encode a partial specification of correct behavior,
have been extensively used to automate software testing for complex systems (e.g.,~\cite{barr_test_oracles, segura2016survey, weyuker1982testing, just2011}. While some partial test oracles specify which properties of a system's output must hold for any input, others specify such properties with respect to a given input or set of inputs.
In our use case, partial test oracles test a property of neural decoding model executions.

Software debugging for traditional software systems (as opposed to learned ML models) involves localizing a fault to its root cause in the program's source code as well as repairing that fault. The software engineering research community has developed a variety of approaches to aid and automate software debugging. For example, statistical fault localization approaches aim to rank program elements by their suspiciousness~\cite{wong2016survey,pearson2017}. Similarly, program slicing approaches~\cite{weiser1984program} aim to identify slices of a program of interest, based on slicing criteria (e.g., derived from data- or control-flow properties.)
Information about a fault's root cause can then be used to repair the faulty system.
The key difference between our application of such debugging techniques and the application to traditional software systems is the focus on data as opposed to programs.

\tableTaxonomy
\section{Classification of BCI Applications and Faults} \label{sec:classification}
Prior work has classified BCIs based on system characteristics and target users and tasks (e.g.,~\cite{1200910, steinert2019doing}). Since differences in BCIs have implications on the ability to define partial test oracles, corrective heuristics, and slice functions, we expand upon existing classifications of BCI applications and additionally classify BCI faults.

\subsection{Classification of BCI Applications} 
\label{sec:bci_classification}
Our classification focuses on four key dimensions along which BCI applications can be classified (Figure \ref{fig:bcitypes}).

\mypara{Prediction Scope}
BCIs vary in their prediction scope. BCIs may predict continuous
outputs (e.g., position and velocity for motor decoding BCIs) or discrete outputs (e.g., stages of sleep).

\mypara{Input Modalities}
BCIs use various types of input modalities (i.e.,
types of neural data analyzed). These range from noninvasive techniques such as magnetoencephalography and electroencephalography (EEG) to invasive but more precise techniques such as intracortical neural recordings and electrocorticography. Despite differences in techniques, most modalities yield time-series data that are occasionally multi-channel. Specific features, such as frequency components of signals or average firing rates, may be computed from raw \mbox{data and used to predict states such as intended movement.}

\mypara{Control Mechanism}
BCIs vary in their control mechanisms (i.e., how a user interacts with the BCI.)
Some motor BCIs require users to actively control their neural patterns through imagination to
signal their intended movement. Other BCIs detect specific types of neural
activity that are evoked upon sensory stimulation. In this way, such BCIs may be able to detect whether the user is selectively focusing on an object, such as a character in a
speech-decoding BCI. Still other BCIs may detect changes in the neural activity that are
not intentionally modulated by the user.

\mypara{Intended Use}
BCIs vary in their intended use. Some are restorative, aiming to provide sensory or
motor function. Others are diagnostic or treatment-centered. Examples of the latter category include deep brain
stimulation, which stimulates deep regions of the brain and has been found to mitigate 
symptoms of Parkinson's Disease and epilepsy, and epilepsy predictors that may alert the
user upon prediction of a seizure~\cite{choi_rhiu_lee_yun_nam_2017}.

\subsection{Classification of BCI Faults} \label{sec:taxonomy}

We classify BCI faults into four non-disjoint categories, summarized in Table~\ref{table:fault_taxonomization}: input validation,
temporal validation, consistency across modalities and models, and domain
knowledge violation.
Categorizing faults in such a way motivates our methodology in which faults are detected through
partial test oracles (Section~\ref{sec:repair_title}).

\mypara{Input validation}
Neural data can be noisy due to muscle movement and electrical interference.
Careful pre-processing and feature extraction to remove the resulting artifacts
in the input data is critical. Since the
inputs that a model sees during training likely have few artifacts due to
controlled experimental conditions, % and careful signal processing, 
artifacts
remaining in the input data due to faulty pre-processing are a particular concern for regular
usage of the BCI after deployment.

\mypara{Temporal validation}
Model output may be temporally incorrect.
In some applications, for example, the 
output class should not change too rapidly between certain states or within a small
range of the output space. Indeed, one of the challenges in developing prosthetics
is producing smooth limb movement when operating in free space or when precisely grasping small objects \cite{6347460}.
Furthermore, certain state transitions are invalid or improbable, due to physiological constraints. Thus, even though
the neural decoder itself may be time-independent and may have no constraints on
the validity between disjoint predictions of the model, temporal faults can occur,
since the model is utilized in a time-dependent context.

\mypara{Consistency}
If present, auxiliary data such as
respiration, heart rate, or movement may be utilized
to enforce consistency between the auxiliary information and the neural decoder's predictions.
Furthermore, some BCIs utilize multiple models or data from different modalities of recording neural signal
to make predictions. For such BCIs, the outputs across these models or modalities must be consistent. For
example, the presence of an artifact in the pre-processed data must be consistent 
across different recording modalities.

\mypara{Domain knowledge}
Domain knowledge about the expected behavior of a
system can aid testing. For example, there may
be physical constraints on limb positions, such as invalid finger positions for a
hand grasping BCI, and there could be patterns in model output that
are known to be improbable based on domain knowledge.
Knowledge of an individual user's
health conditions and daily activities may also be applied to enforce constraints.

\section{Repairing BCIs with Fault-Based Data Acquisition} \label{sec:repair_title}

This section details the proposed methodology depicted in Figure~\ref{fig:workflow}. Sections~\ref{sec:identifying}--\ref{sec:localization} describe partial test oracles, corrective heuristics, and slice functions, and section~\ref{sec:repair} details how these components are used for targeted, fault-based data acquisition and repair.

\subsection{Detecting Faults with Partial Test Oracles} \label{sec:identifying}

Our methodology uses partial
test oracles as a means to test BCIs and detect faulty outputs
of the underlying neural decoding model.
The behavior and expected output of a neural decoder usually cannot
be precisely specified, which makes partial oracles especially suitable for testing. These oracles capture domain knowledge of a partial
specification of a BCI---specifically, what properties
of the neural decoder output violate necessary conditions.

\mypara{Partial Test Oracles for Input Validation}
Software and hardware systems usually
have defined preconditions that must be met for valid execution.
In software systems for example, these preconditions can be expressed through assertions.
In the context of BCIs, input validation is similar to such preconditions:
if pre-processed neural data is dissimilar to the type of data a decoder
was trained on, the decoder cannot be expected to produce valid output. Artifacts
that occur in neural data due to unrelated movement and interference with the neural data
collection are often removed by signal processing before the data is passed
through the neural decoder. Partial test oracles can be defined to capture cases in which these
artifacts do not get removed from the data input to the decoder.
These cases can then be
later analyzed and used to either improve the signal processing steps or improve
the generalizability and robustness of the decoder.

\mypara{Partial Test Oracles for Temporal Validation}
Partial test oracles can also capture neural decoder faults that are temporal.
Oracles for temporal faults
utilize ideas from metamorphic testing in that previous
model output is considered when testing subsequent model output. Metamorphic
relations are useful, for example, for detecting illegal transitions in model
output.

\mypara{Partial Test Oracles for Consistency}
Partial test oracles can detect inconsistencies between different models or across different modalities of neural recording.
For example, a simpler, precise auxiliary model that
accepts the same input as a neural decoder, or related data,
can be used to identify faults in the neural decoder's output.

\mypara{Partial Test Oracles for Domain Knowledge}
Test oracles can also capture faults based on domain knowledge of the
likelihood of various events.
For example, clinicians may have noticed in experiments
that a certain state transition in the model output space is extremely unlikely in natural execution or
that a particular position is physically impossible or unsafe. Such experience-based constraints
can be defined in form of a partial test oracle.

\subsection{Transforming Implausible to Plausible Model Outputs with Corrective Heuristics} \label{sec:heuristics}

While partial test oracles can detect faulty predictions, correcting these predictions is challenging because the correct output is unknown---as is the user's activity during regular usage of the BCI that resulted in the fault occurring.

Our methodology uses corrective heuristics that transform faulty predictions by mapping implausible output values to plausible ones. A different corrective heuristic is defined for each type of fault.
For example, a corrective heuristic may transform an illegal transition
by changing the faulty prediction to the previously predicted state or to a state that is reachable from the previous state.
Similarly, a corrective heuristic may transform an out-of-bounds coordinate to the nearest in-bound coordinate. While there is no guarantee that the resulting value is itself correct, it is at least plausible.

Corrective heuristics may utilize auxiliary information, such as a
state transition probability matrix learned on a ground truth dataset, while others
may generate corrections based on the neural decoder's predictions on surrounding data points.
Corrective heuristics for neural decoders are specific to a fault in a specific BCI application, but they do share some common characteristics.
First, unlike labeling functions used for weak supervision
\cite{ratner2020snorkel}, corrective heuristics only apply to faulty predictions.
Second, since BCIs operate on time-series data (i.e., neural signals over time), corrective heuristics rely on previous predictions. For example, a corrective heuristic for illegal state transitions may assume that the previous decoder predictions were correct and may select the most common recently predicted state as the plausible output.

While corrective heuristics can transform individual faulty predictions, they do not truly repair the neural decoding model itself, and hence cannot be expected to generalize to long-term BCI usage. Consequently, our methodology uses corrective heuristics to improve fault localization but not as a standalone solution to repair.

\subsection{Localizing Faults with Slice Functions} \label{sec:localization}
Our methodology uses slice functions to localize a detected fault to co-occurring input slices and output slices. Guided by domain knowledge, these slice functions capture characteristics of the input and map outputs to tasks, given a stream of neural decoder model executions. For applications with discrete output, output classes can be directly mapped to tasks; for applications with continuous output, \mbox{the output is discretized into bins, which correspond to tasks.}

While output slices are directly related to tasks that the user might perform during daily usage of the BCI, input slices are related to subtler characteristics such as data artifacts, environment settings, or characteristics of the user. Localizing faults to input slices enables possible refinement of tasks (e.g., performing a task in a particular environment setting) and improvement of BCI components such as pre-processing (e.g., artifact removal).

This paper studies the effectiveness of localizing faults to both input and output slices but only uses the output slices (tasks) to guide targeted data acquisition by generating a fault-based task distribution. We conjecture that faults localized to input slices can inform BCI developers about necessary improvements of data-processing components. We leave a deeper exploration as future work.

\subsection{\fontsize{10.5pt}{11pt}\selectfont\mbox{Retraining with Fault-Based Data Acquisition}} \label{sec:repair}
Our methodology uses fault-based data acquisition that is focused on eliminating specific faults. Simply retraining a neural decoder on heuristically corrected data or randomly sampling data is insufficient.
First, heuristically corrected data, when directly used for retraining, would be inherently noisy. Second, non-targeted data acquisition is inefficient because faults are not uniformly distributed across tasks. Targeted, fault-based data acquisition, by contrast, yields a more focused training dataset.

\tableApps

Data acquisition is the standard method for training and recalibrating BCIs for users' needs. However, the traditional approach to data acquisition is not targeted at eliminating particular faults. For example, a user's report that the BCI sometimes rapidly switches between acquiring and releasing more than once for a single grasping task has an unknown root cause and cannot be easily addressed.

Using partial test oracles, corrective heuristics, and slice functions, our methodology computes a coincidence distribution of tasks based on fault occurrences from past BCI usage. This computed distribution of tasks may indicate that a particular task (e.g., a stage of acquiring or grasping) tends to coincide with faults more often and thus requires more training data to inform the model's predictions for that faulty task. While the example of simple grasp control has a small set of tasks, BCIs that control a wider range of hand gestures perform a larger set of tasks. The more tasks a BCI performs, the more important it is to acquire training data according \mbox{to a task distribution to effectively target and eliminate faults.}

The output slice functions are applied to the heuristically corrected model outputs to determine a more accurate task distribution (e.g., an illegal state or out-of-bounds prediction cannot be directly mapped to a task).
The task distribution is then used to retrain the underlying model.

\section{Evaluation Approach}
\label{sec:apps}

Section~\ref{sec:repair_title} proposes a methodology for detecting and repairing BCI faults. To evaluate the efficacy of the individual components of this methodology, we studied the following three research questions:

\begin{itemize}
  \item \rqOne
  \item \rqTwo
  \item \rqThree
\end{itemize}

\noindent
To evaluate the efficacy of BCI repair via targeted, fault-based data acquisition, we answer the following two research questions:

\begin{itemize}
  \item \rqFour
  \item \rqFive
\end{itemize}

To answer our research questions, we selected 5 different
neural decoding BCI applications (summarized in Table~\ref{table:decoders}).
Each decoder uses neural data (EEG or single neuron recordings) in some
model species (human, monkey, or rat) to decode an aspect of behavior (motion or sleep). We
used publicly available labeled datasets and existing implementations of neural decoding models.

We split each ground truth labeled dataset into a training set for
initial training and calibration, an observation 
set to simulate regular usage, an acquisition set that 
can be sampled from to simulate clinical data acquisition,
and a testing set for evaluating performance.
We generally split each dataset based on a 6:2:1:1 ratio. For the Field Navigation application, we observed during initial exploration that 20\% of the data had substantially lower quality than the rest; we discarded the data. The baseline model of the Cursor Control application had a fault frequency very close to zero, which would have rendered it as not suitable for our evaluation. We discarded 20\% of the training data, still leaving the Cursor Control application with a substantially lower fault frequency than all other applications (Table~\ref{table:prevalence}).

We first trained each neural decoder on the training set to achieve a
baseline model. We then made predictions on the observation set using the baseline
neural decoder, simulating regular usage of the BCI and capturing fault occurrences. We used the
collected faults to perform some method of repair and re-assess the performance. The model
performance was always computed on the testing set. For the results on performance and fault frequency,
we ran ten trials using different splits of the dataset and averaged the results.

\tableHeuristics

\subsection{Selected BCI Applications}
We selected five neural decoder applications (Table~\ref{table:decoders}), aiming at a representative selection based on the classification in Section~\ref{sec:bci_classification}. Specifically, we selected both discrete-output and continuous-output decoding BCI applications, those that use either EEG or single-neuron recording (SNR) as the input modality, both active and passive BCIs, and both restorative and diagnostic BCIs.
Selecting BCIs that vary along the four dimensions enables us to more comprehensively evaluate our proposed methodology.
For example, BCIs with discrete output classes allow for testing output validity based on state transition violation, while those with continuous output space require assertions on the magnitude and rate of change of the output and also make defining accurate corrective  heuristics more challenging.
Furthermore, classical ML models usually operate on manually engineered features, while deep neural networks do not require such feature engineering, and this may have implications on the ability to repair faults.

\subsubsection{EEG Motor Decoding (MD)}
We used the EEG Motor Movement/Imagery Dataset from Physionet to
train an LSTM-based neural decoder \cite{goldberger_2000,schalk_mcfarland_hinterberger_birbaumer_wolpaw_2004,zhang2020survey} to classify three fist positions. EEG data was
recorded from 64 electrodes placed in precise locations on the scalp
while participants closed their left fist, right fist, or 
neither based on the location of a cursor on a screen. This decoder has immediate applications in a hand prosthetic that 
must decode the user's intention to open or close a particular fist to grasp an object.

\subsubsection{Observation-or-Execution Classification (OE)}
In this application, a support vector machine (SVM) model was used to classify
between action observation and motor execution of the sit to stand action \cite{chaisaen2020, Chaisaen_2020_decoding}. EEG and electromyography (EMG) data were collected while participants were shown a video stimulus of someone performing the
sitting to standing task (action observation), given a rest period, and then asked
to execute that task (motor execution).
The data collected while executing the action was processed to extract movement related cortical potentials before being classified. This decoder has applications in motor
prosthesis, because distinguishing the intent of a user to perform an action vs. a
passive thought or observation of an action is critical for correctness.

\subsubsection{Sleep Stage Classification (SS)}
For application diversity, we selected a passive (diagnostic) BCI which classifies sleep/wake stages into five possible outcomes.
We used the Sleep-EDF dataset containing whole-night PolySomnoGraphic EEG recordings and a Random
Forest Classifier to classify between stages \cite{GramfortEtAl2013a, chambon2018}. Such a BCI can help monitor and detect abnormalities 
in individuals sleep patterns that could signify underlying medical conditions.

\subsubsection{Cursor Control (CC)}
In this continuous decoding application, a Wiener Cascade decoder
predicts the x- and y- coordinates of a cursor on a 2D screen controlled 
by a monkey physically manipulating a manipulandum. We used a dataset
that contained single-cell recordings from 52 neurons in the brain and was acquired from the 
Kording Lab \cite{Glaser2020}. The ability to control a digital cursor 
with one's mind has significant implications for individuals with 
impaired mobility and communication skills.

\subsubsection{Field Navigation (FN)}
The final application involved using a deep neural network to predict
the x- and y-coordinates of a rat navigating towards rewards on a
platform \cite{Glaser2020}. We utilized another dataset from the Kording
Lab, which contains single-cell recordings from the rat's hippocampus, a part of the brain involved in memory. Such motor
movement-related BCI applications have practical applications for developing prostheses.

\subsection{Partial Test Oracles} \label{sec:oracles}

We defined partial test oracles
for each of the five BCI applications.
Most of these oracles are stateful (i.e., state is maintained across multiple executions of the decoder model) and are similar to metamorphic relations~\cite{segura2016survey}. Given each new model execution, a partial test oracle specifies what executions among the current and past consecutive model executions are faulty, if any.

\subsubsection{EEG Motor Decoding (MD)}
In this application, we identified \textit{illegal state
transition} and \textit{temporal inconsistency} faults, both of which fall under
the category of temporal validation. First, based on the experimental setup for
this application, it is impossible for the participant to directly switch between
having their left fist closed and right fist closed, and thus transitions between
these states can be considered ``illegal.'' Furthermore, due to the presence of noise
in the input EEG signals and limitations of the baseline decoder, the model's
prediction may flicker rapidly between states over a period of time instead of
consistently remaining in a certain state for a reasonable amount of time
(\textit{temporal inconsistency}). 

\subsubsection{Observation-or-Execution Classification (OE)}
For the observe-or-execute classification application, we focused on a
temporal inconsistency fault: an individual cannot rapidly switch between action
observation and motor
execution since they remain in each state for a non-trivial period of time, and thus the output
of the decoder should not flicker between those states.

\subsubsection{Sleep Stage Classification (SS)}
For the sleep stage classification application, we cover three faults, of which two
fall into the category of temporal validation. We identified an \textit{illegal
transition} fault as it is physiologically impossible to transition between certain
stages of sleep, and a \textit{temporal inconsistency} fault because individuals
remain in one stage of sleep for an extended period of time. The third partial test oracle (for the \textit{multimodal inconsistency} fault) identified inconsistencies between the neural decoder's output and the output of a
binary classifier that classifies whether the BCI user is awake or sleeping
based on auxiliary data, including respiration, body temperature, and EMG data.

\subsubsection{Cursor Control (CC)}
For the cursor control application, we identified two different temporal faults, temporal inconsistency and rapid motion. 
First, the BCI can make multiple faulty
predictions of position values that change significantly in a short period of time (\textit{temporal inconsistency}) when the cursor movement should be smooth. Second, the BCI can consecutively
predict two position values whose distance exceeds a given threshold (\textit{rapid motion}), when the cursor should move at a slower pace.

\subsubsection{Field Navigation (FN)}
For the field navigation application, we defined three faults. 
First, the BCI can make an \textit{out of bounds} prediction,
incorrectly predicting a position coordinate that lies outside the field. This is an example
of a violation of domain knowledge. Second, we detected \textit{temporal
inconsistency} faults in this application: the position in the field cannot
change too rapidly due to physical limitations to rat movement.
Third, we defined the \textit{rapid motion} fault as predictions with unreasonable sudden jumps.

\tablePrevalence

\subsection{Corrective Heuristics and Slice Functions} \label{sec:description_functions}
We defined corrective heuristics for each of the five BCI applications. Table~\ref{table:faults} summarizes these heuristics and the faults they apply to.

For each of the five BCI applications, 
we defined slice functions for capturing discrete input and output slices, given a stream of neural decoder model executions.
For the three applications with discrete output (motor decoding, observe-or-execute classification, and sleep stage classification), we defined output slice functions that map output classes to tasks.
For both the Cursor Control and the Field Navigation application, we defined output slice functions to classify the direction of movement of the cursor or the rat, respectively.
For the Cursor Control application, we additionally defined output slice functions to classify the position (mapping each coordinate to a quadrant subspace of the screen) of the cursor.

Both the Cursor
Control application and the Field Navigation application additionally utilized functions that capture input slices. We defined two slices based on properties of the firing 
activity of single neurons. The \textit{hypoactivity} and \textit{hyperactivity} slice functions detect when the neural signal over an entire time bin is below or above certain thresholds, respectively.

\subsection{Retraining and Evaluating Repair}
While the Motor Decoding and
Field Navigation neural decoders used neural networks
and could be efficiently retrained by incrementally training on new data, the other three decoders used traditional ML models and required retraining with a concatenation of the initial and new training data.
To acquire the new training data, we sampled 500 data points from the acquisition dataset.

To assess the effectiveness of our methodology for repairing BCIs, we compared both fault frequency and overall prediction performance of a repaired model to the baseline model. We specifically assess whether different data acquisition strategies lead to significant differences.

\section{Evaluation Results}
\label{sec:results}

\subsection{\rqOne}

Since faults are detected through partial test oracles, we computed precision as the percentage of correctly detected faults, based on ground truth, out of all faults detected by an oracle.

\mypara{Precision of BCI Fault Detection via Partial Test Oracles}
According to
Table~\ref{table:prevalence}, the oracle capturing inconsistency with
the auxiliary model (Multimodal Inconsistency) has the greatest precision, followed by the
oracles capturing illegal transitions,
as they identify states of the decoder
that are certainly incorrect. While the other oracles have lower precision, they
can be fine-tuned to improve precision. A test oracle that detects temporal
inconsistency, for example, can be hard to define, as one must fine-tune the
window size of model executions to consider at a time to determine inconsistency.
However, we observe in Section~\ref{sec:RQ5} that
oracles with lower precision are still valuable in improving decoder performance.

\mypara{Prevalence of BCI Faults}
According to Table~\ref{table:prevalence}, the most prevalent faults detected
by partial test oracles across all the
applications are illegal transition and temporal inconsistency. This is not
surprising because neural decoders process temporal data that varies in very specific
ways across time. Furthermore, the prevalence of all faults indicates that 
there is potential for improving the performance of neural decoders by eliminating faults.

\subsection{\rqTwo} \label{sec:RQ2}

We evaluated the efficacy of each corrective heuristics, based on ground truth labels. Specifically, we compared model accuracy before and after applying the corrective heuristics to the baseline neural decoder predictions. Note that the performance for the cursor control and field navigation applications is measured using mean squared error, for which lower is better. The baseline is the faulty neural decoding model that was initially trained for regular BCI usage.
Figure~\ref{fig:decoder_correction} gives the results and shows that
4 out of 5 BCI applications demonstrated a significant improvement in accuracy after applying the corrective heuristics. The accuracy of the motor decoding model decreased, but the difference was not significant.

\subsection{\rqThree} \label{sec:RQ3}

\figureHeuristics
\figureLocalization

In order to test whether faults can be localized to input and output slices, we compared the distributions of input and output slices, conditioned on whether a particular type of fault occurred. We used a Chi-squared test of independence on the relationship between the presence of a given type of fault and the distribution of input and output slices, respectively.

Figure~\ref{fig:fault_task} shows that most types of faults exhibit distinct distributions of input and/or output slices for---both compared to the distributions of other faults and no fault in a given application.
For example, temporal inconsistency faults in the motor decoding application coincide with instances of the resting task, whereas illegal transition faults almost never occur during resting (Chi-squared test of independence, p<0.05). In this case, the illegal transition fault can be localized to a task distribution primarily comprised of the non-resting tasks.
The same observation holds for the observe-or-execute classification, with only one type of fault. The distribution of tasks being simultaneously performed during fault occurrence is significantly different than the no-fault distribution (Chi-squared test of independence, p<0.05): when temporal inconsistency faults occur,
a BCI user is more likely to be observing an action. In this case, temporal inconsistency faults can be localized to a distribution skewed towards the observation task.

Overall, our results show that for most types of faults in the five BCI applications, there is a statistically significant association between fault occurrence and slice distribution.

\vspace*{-4pt}
\subsection{\rqFour} \label{sec:RQ4}

To evaluate our proposed methodology for fault-based data acquisition and repair, we trained the underlying neural decoder on data acquired based on the computed fault-based task distributions, for all five BCI applications. We measured repair success for each type of fault in isolation as well as for all types of faults combined. Specifically, we computed the change in the number of fault occurrences after training. Although individual types of faults can co-occur, each fault represents a unique issue. Therefore, when measuring repair success for all types of faults combined, we sum the fault occurrences of each individual type of fault (as opposed to counting unique faulty model outputs).

Table~\ref{table:repair_loc} gives the overall results and shows that retraining the underlying neural decoding model after fault-based data acquisition can reduce the number of faults.
For example, for the observe-or-execute classification application, training on data acquired through a task distribution yields a statistically significant improvement over the baseline and over training on data acquired through the natural task distribution.

For other applications, we did not observe such a
statistical significant reduction, and we conjecture three reasons.
First, the acquisition dataset is limited, being only one-tenth of the original dataset. Since we acquired data by selecting a fixed number of datapoints from this already limited acquisition dataset, there is less room for the task distribution to show a significant effect, compared to the natural distribution.
Second, since we use previously collected data sets, we cannot acquire new data based on the task distribution, but instead simulate data acquisition by sampling from a portion of the existing dataset that the model has never seen. Fault-based data acquisition in a real calibration session may be more targeted.
Third, the corrective heuristics that we defined for this evaluation may not be accurate enough to produce accurate task distributions.
We wish to explore these further in future work, and carrying
out a user study may provide additional insights.

\tableRepair

Fault-based data acquisition can achieve a reduction in faults. However, a secondary concern is that retraining must simultaneously ensure that accuracy (or related measure of performance such as mean-squared error) does not significantly degrade. We evaluate this by computing the change in accuracy after retraining on data acquired through fault localization.
We found that there was a significant \emph{increase} in accuracy for the motor decoding application, and no significant change for the other applications, suggesting that fault-based \mbox{data acquisition can repair BCIs without degrading accuracy.}

\subsection{\rqFive} \label{sec:RQ5}
Section~\ref{sec:RQ2} established the effectiveness of fault localization based on heuristically corrected model outputs, used to compute a task distribution for data acquisition and model retraining.
This method requires both (1) corrective heuristics to correct faulty predictions and (2) fault localization to inform data acquisition. Here, we explore whether either of these components suffices. In other words, we test whether either training on heuristically corrected predictions without a fault-based distribution or localizing faults to tasks without correcting faulty predictions would confer performance improvements.
The two right-most columns in Table~\ref{table:repair_loc} show the results for these two alternative approaches.

First, we discuss
the results of directly retraining the neural decoder on heuristically corrected faults, without assuming a fault-based task distribution.
As shown in Table~\ref{table:repair_loc}, we observe that training on heuristically corrected faults decreased the number of faults for the motor decoding application. However, we also observed a significant increase in the number of faults for 3 out of 5 applications.
We attribute this to two reasons.
First, the amount of the data the decoder model is trained on is dependent on the number of detected faults. In the applications
that we tested, fault frequency is overall low, and training on this
small number of specific data points likely causes the model to overfit.
Second, since corrections are based on heuristics, training on them could bias the model, especially if there are few corrected datapoints to retrain on. We conclude that while corrective
heuristics are a useful tool for correcting individual predictions,
training on the corrected data points is insufficient for repairing the
underlying decoder.

Next, we discuss the effectiveness of fault-based data acquisition without the use of corrective heuristics. In Section~\ref{sec:RQ4}, corrective heuristics were used as a means for weakly correcting model predictions in order to accurately determine the task being performed. However, correcting model predictions with imprecise heuristics may 
not result in a more accurate task distribution.
To test whether effective fault-localization could be achieved without corrective heuristics, we computed
the number of faults when retraining on fault-based acquired data but without applying corrective heuristics to model predictions first.
We found that removing corrective heuristics significantly decreased the number of faults only for illegal-transition faults in the sleep stage classification application. For all other faults and applications, the results of this approach are comparable to or significantly worse than the results of fault-based data acquisition with corrective heuristics applied.
We conclude that corrective heuristics are an important component in the methodology to compute an accurate fault-based task distribution.

\section{Threats to Validity}

\mypara{External Validity}
We selected five BCI applications for evaluating the
applicability of our proposed methodology for repairing BCIs post-deployment.
The comprehensiveness and representativeness of this
selection affects external validity, but we aimed for a diverse set of applications by considering the different dimensions for classifying BCIs (Section~\ref{sec:classification}).
Additionally, our methodology assumes that neural decoder model predictions can be mapped to a sufficiently large number of tasks for fine-grained fault localization. We argue that this assumption is satisfied in practice: BCIs such as neuroprostheses indeed perform a variety of tasks. For example, hand posture can be classified into more than 40 types, and since BCIs are an emerging technology, advances will likely results in an even greater number of supported postures.

\mypara{Construct Validity}
Our evaluation used existing, labeled datasets to simulate real-world BCI usage. These datasets were collected in a rigid and non-naturalistic experiment settings. For example, in the EEG motor decoding application, participants opened and closed their fists on cue, but during real-world usage these actions would be naturalistic. As a result, our efficacy measures of partial oracles, corrective heuristics, and slice functions could be specific to the non-naturalistic setting.
A related concern is that our measures for fault occurrence and frequency may be specific to the experimental setting and are not accurate measures for faults that will occur during real-world usage.
Likewise, faults that \textit{do} appear in an experimental setting may not occur during real-world usage. However, many of these faults such
as temporal inconsistency do occur in other ML-based control systems~\cite{Kang2018}. Thus, while most BCI research has reported on overall model performance (as opposed to reporting on fault occurrences), it is reasonable to expect that these faults do occur in BCIs as well.

\mypara{Internal Validity}
A threat to internal validity is model overfitting, which could prevent our evaluation from properly isolating the effect of retraining based on a given task distribution. Recall that our evaluation used existing data sets and retrained models on small, fixed sets of newly acquired data. There is a risk of overfitting to the initial training data and/or to the newly acquired data, which could worsen model performance. This, however, would likely underestimate the efficacy of fault-based data acquisition because real-world data acquisition would not be subject to the same restrictions.

\section{Related Work}

There are three avenues of closely related work: (1) methods for improving the 
performance of BCIs, (2) applications of software testing research to machine learning 
systems, and (3) miscellaneous work combining BCIs and software engineering.

\mypara{Optimizing BCI Performance}
Prior work has sought to improve the data collection methodology for BCIs through
emerging technologies or novel sampling techniques.
For example, virtual reality has been applied to more realistically simulate
the surrounding environment of a BCI user \cite{boukhalfi2015tools}. Active class selection (ACS) has also been previously explored as a means for more 
efficient
data acquisition for BCIs. A combination of entropy and the inverse of class accuracy has been
applied to motor imagery for choosing classes that are informative and in need of improved 
accuracy \cite{hossain2017weighted}. Data acquisition through ACS has been combined with
transfer learning in an application to a BCI \cite{wu2013collaborative}. By contrast,
instead of the general goal of increasing accuracy of neural decoding predictions,
our stated goal is to repair specific faults as reported by users. Additionally instead
of applying a sophisticated sampling algorithm, we opt for a more
software-testing-oriented and human-in-the-loop approach with partial test oracles defined by domain experts.

\mypara{Software Testing and ML Systems}
The problems encountered when testing BCIs are similar to those encountered when testing other data-driven ML systems. For example, prior work has shown that neural network accuracy can be improved through fault localization by localizing errors to specific neurons in the network 
\cite{eniser2019deepfault} and that fairness and accuracy can be improved by localizing errors to influential data points~\cite{koh2017,verma2021}.
Machine learning has also been applied to improve debugging techniques for traditional software systems (e.g.,~\cite{nath2016learning, ascari2009exploring, li2019deepfl}).

\mypara{Software Engineering and BCIs}
Other work has sought to apply software
engineering to BCIs. Application areas have included
writing, comprehension, and debugging of code
\cite{boukhalfi2015tools, duraes2016wap, peitek2018look}. The application of software 
engineering research to BCIs has been proposed 
with a particular emphasis on self-adaptive systems reacting to inputs where those inputs are 
now neural signals \cite{huang2014brainware}.

\section{Conclusion}

This paper explores the application of software testing and debugging methods to the emerging field of brain-computer interfaces (BCIs). Specifically, it proposes a human-in-the-loop methodology for testing and repairing BCIs post-deployment.
The evaluation based on five BCI applications shows that (1) partial test oracles can indeed detect faults in BCIs, (2) corrective heuristics and slice functions can localize faults to specific data slices in both the input and output spaces, and (3) fault-based data acquisition can reduce the fault frequency while maintaining model accuracy.
Overall, the results suggest that the proposed methodology is a promising step towards more accurate and safer BCIs.

\bibliographystyle{ACM-Reference-Format}
\bibliography{paper}

\end{document}